\documentclass[twocolumn,showpacs,preprintnumbers,amssymb]{revtex4}
\usepackage{graphicx}
\usepackage{dcolumn}
\usepackage{bm}
\begin{document}
\preprint{CCOC-03-100 (Last Update:\today)}
\title{Characteristics of a Delayed System with Time-dependent Delay Time } 
\author{Won-Ho Kye} 
\email{whkyes@empal.com}
\author{Muhan Choi}
\author{Sunghwan Rim}
\author{M.S. Kurdoglyan}
\author{Chil-Min Kim}
\email{chmkim@mail.paichai.ac.kr}
\affiliation{National Creative Research Initiative Center for Controlling Optical Chaos,
Pai-Chai University, Daejeon 302-735, Korea}
\author{Young-Jai Park}
\email{yjpark@ccs.sogang.ac.kr}
\affiliation{Department of Physics, Sogang University, Seoul 121-742, Korea}
\begin{abstract}
The characteristics of a time-delayed system with time-dependent delay time is investigated. 
We demonstrate the nonlinearity characteristics of the time-delayed system are significantly 
changed depending on the properties of time-dependent delay time and   
especially that the reconstructed phase trajectory of the system 
is not collapsed into simple manifold, differently from the delayed system with fixed delay time.
We discuss the possibility of a phase space reconstruction and its applications.
\end{abstract}

\pacs{05.45.Xt, 05.40.Pq}
\maketitle
The effect of time delay due to a finite propagation speed 
of information is usually considered as the form of delay-differential equation: 
$\dot{x}=f(x(t), x(t-\tau_0))$, where $\tau_0$ is the fixed delay 
time \cite{Opt,MG,Bio,Chem,Cont,Comm,Measure,Farmer,TD_Sync,Map_Delay,TD_Recon,Bunner}.
It has been found that the system actually exhibits many different behaviors depending 
on the nonlinearity and the delay of the system
and that the dimension of the attractor rises linearly with the delay time, even though
the number of degree of freedom is small \cite{Farmer,TD_Sync,Map_Delay,TD_Recon}. 
In the last decade, models based of delay time have been 
extensively investigated
in various fields such as optics\cite{Opt}, biology \cite{MG,Bio} and chemistry \cite{Chem}
for the purpose of understanding its fundamental role 
and of applying it to control \cite{Cont} and communication \cite{Comm,Measure}. 

The models based on fixed delay time, however,
often fail to properly cover such real factors as (a) memory effect of the oscillator, 
(b) approximately known delay time, and (c) time-dependent delay time \cite{Vol,DDT1,DDT2}.
To cover these factors, Volterra first proposed a model based on distributed delays \cite{Vol}.
The model has been used in various areas \cite{DDT1,DDT2,Vol_Area}. 
It has been shown very recently that the distributed delay induces 
a death phenomenon in a much larger set of parameters than that of the fixed delay \cite{DDT2}.
Thus the Volterra's model has enabled us to understand the realistic 
effects of delay times in dynamical systems.  

Meanwhile, in studying the population dynamics and epidemic problems the delay time 
has been considered as a function of state variable \cite{Hale} and there have been
extensive investigations in that direction. 
However, there are many real situations in which the dynamics 
of delay time can not be described by an analytic function, e.g., neural networks and internet \cite{DynTopo}.  
So it is reasonable to introduce  
time-dependent delay time as a stochastic process in those cases. 
In this point of view we shall investigate the effects 
of time-dependent delay time (TDT) in dynamical
systems governed by a stochastic process 
and the effects in time-delayed systems remain much less studied. 
The main goal of this paper is to show how TDT alters the characteristics 
of time-delayed systems. In addition, we analyze these characteristics  
with regards to application to communication. 
%
\begin{figure}
\begin{center}
\rotatebox[origin=c]{0}{\includegraphics[width=8.3cm]{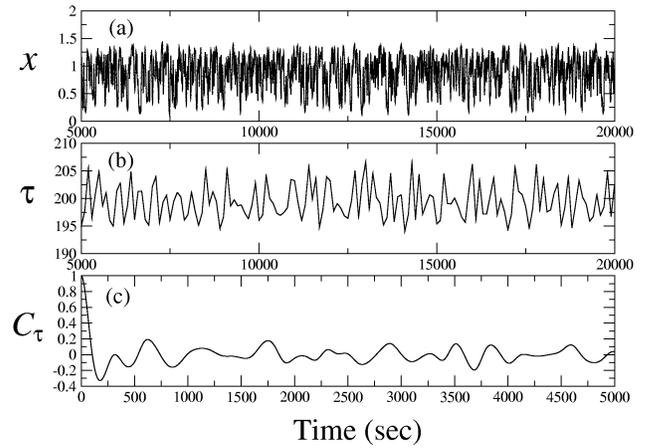}}
\caption{ temporal behaviors of the modified Mackey-Glass model in the presence of TDT when $\tau_0=200$, $T=100$, and $\Lambda=10$:
        (a) state variable $x$; (b) the delay time; (c) autocorrelation of the time-dependent delay time. }
\end{center}
\end{figure}
We consider the modified Mackey-Glass model. The Mackey-Glass model \cite{MG} was introduced 
as a model showing the regeneration of blood cells
in patients suffering from leukemia. 
The modified Mackey-Glass model is given by,
\begin{eqnarray}
\dot{x} &=& \frac{a x(t-\tau(t))}{1+x^{10}(t-\tau(t))} -b x(t), \nonumber\\ 
{\tau}(t) &= &\tau_0+\int ^{t}_0\xi(s) ds,
\end{eqnarray} 
where $a=0.2$, $b=0.1$.
While in the Mackey-Glass model the delay time is a constant $\tau= \tau_0$, 
in the modified Mackey-Glass model $\tau$ is a function of time.
Especially we focus on the case where  $\tau(t)$ is governed by a stochastic process $\xi(t)$.
As an example, we introduce the signal $\bar{\xi}(t)$ 
which is generated by the discrete sampling of the chaotic signal $x(t)$ such that
$\bar{\xi}(t) =  \frac{{x}(nT)-{x}((n-1)T)}{T}\Lambda$, 
when $t \in [nT, (n+1)T)$. 
We note that 
this form of the signal was taken for the convenience's sake, which
allows us to adjust the correlation length and modulation amplitude.
And $\bar{\xi}$ actually exhibits quasi-stochastic signal because 
we shall study the sampling period of $T\in [100, 1500]$  
which is the larger than the correlation length of $x$ (the correlation length of x is $\tau_x\approx 70$ 
in the same parameters of Fig. 1.
Our main results would not be changed if we use real
stochastic signal for driving the delay time,
because those results are related with the fact that the
delay time is not determinable).
Here, $\Lambda$ and $T$ are control parameters for the stochastic signal. 
They are proportional
to the amplitude of $\tau(t)$ and its correlation length, respectively. 
The limit $\Lambda\rightarrow 0$ restores the system to the Mackey-Glass model.
Figure 1 (a) and (b) show the temporal behavior of the modified Mackey-Glass model with TDT.
In Fig. 1 (c), one see that the $\tau$ has the the correlation length of $O(T)$.   



One of the most sensitive measures  
to detect the delay time of a system \cite{Measure,TD_Recon} is the one step prediction error \cite{Bunner}.
In the sufficiently small patch $U_j$ on $(x_i, x_{i-\tau})$ plane, it is defined by
$\sigma_j^2(\tau)=\frac{1}{N_{U_j}} \sum_{\vec{v}_i \in U_j} (\hat{\dot{x}}_i -g_j(\vec{v}_i))^2$,
where $\vec{v}_i=(x_i, x_{i-\tau})$ and $N_{U_j}$ are the numbers of data points in patch $U_j$.
Here, $\hat{\dot{x}}=\frac{x(t+\delta t)-x(t)}{\delta t}$ is the time variation 
obtained from the observed signal $x(t)$. $\delta t$ is the sampling interval 
which should be much less than the characteristic time scale of the system 
(we took $\delta t=10^{-3}$).  
$g_j$ is a local linear function such that $g_j(\vec{v}_j)=b_j + \vec{a}\cdot \vec{v_j}$, 
where the parameters $b_j$ and $\vec{a}_j$ are
determined by the least square fitting. When this occurs, $\sigma_j$ is minimized \cite{Coeff}. 
Therefore, the one step prediction error $\sigma$ is the average of the minimized $\sigma_j$, i.e.,
$\sigma=\langle \sigma_j \rangle$.
\begin{figure}
\begin{center}
\rotatebox[origin=c]{0}{\includegraphics[width=8.3cm]{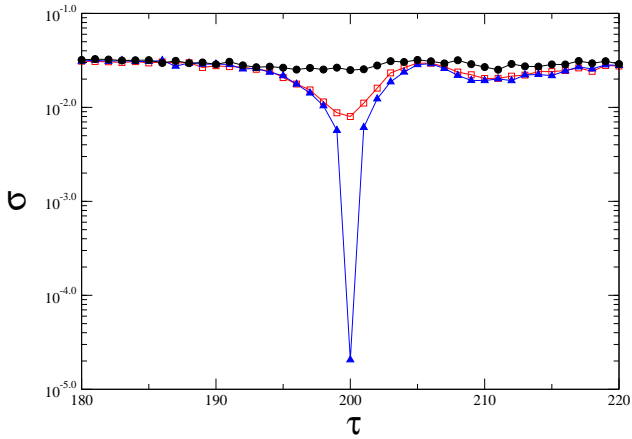}}
\caption{One step prediction errors for $\Lambda=0$ (filled triangle), $\Lambda=10$ (square), and
$\Lambda=50$ (filled circle).}
\end{center}
\end{figure}

Figure 2 shows the one step prediction errors 
for the modified Mackey-Glass model depending on $\Lambda$.
In the case of a fixed delay time (filled triangles) one can see that
$\sigma$(which has a value of $\sigma=1.8\times 10^{-5}$) has a sharp peak at $\tau=200$.
However, if the time dependency of delay time is turned on, i.e., $\Lambda \neq 0$,
the depth of the peak decreases as the $\Lambda$ increases. 
Eventually, the peak almost disappears.  At $\Lambda=50$ (filled circles) 
the prediction error has an almost constant value of $3.2 \times 10^{-2}$.
This means that if one detects the fixed delay time $\tau_0$, 
one can predict the time series $10^{3}$ times as precisely. 
As we shall see, it is closely related to the fact that the phase trajectory on 
the $(\dot{x}(t), x(t), x(t-\tau))$ space collapsed into
the simple manifold.
This is the crucial feature of the system based on the fixed delay time $\tau_{0}$.
And B\"unner et al. \cite{Bunner} have shown that the delayed system can be modeled 
by the time delay embedding when $\tau_0$ is exactly detected.  
Meanwhile, in the case of TDT 
any detectable imprint is not left in the prediction error
above the appropriate value of $\Lambda$.
This feature indicates that the conventional modeling 
method for the delayed systems is not directly applicable.  

\begin{figure}
\begin{center}
\rotatebox[origin=c]{0}{\includegraphics[width=8.3cm]{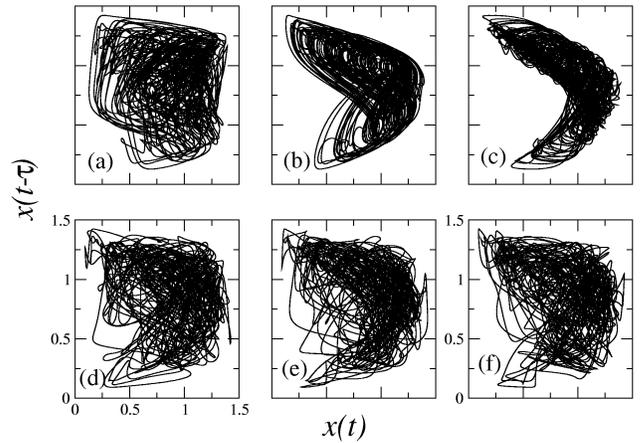}}
\caption{Reconstructed phase trajectories of the modified Mackey-Glass model in 
$(x(t), x(t-\tau))$ space:(a)-(c) fixed delay time ($\Lambda=0$); 
(d)-(f) time-dependent delay time ($\Lambda=50$).
The reconstructions were performed at the lag time $\tau=190$((a) and (d)), 
$200$((b) and (e)), and $210$((c) and (f)), respectively.}
\end{center}
\end{figure}

\begin{figure}
\begin{center}
\rotatebox[origin=c]{0}{\includegraphics[width=8.3cm]{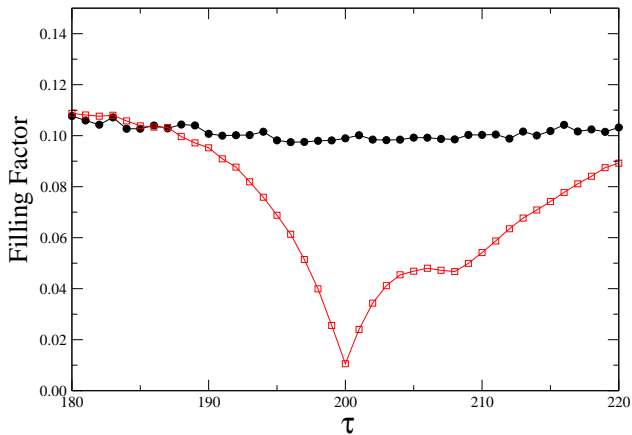}}
\caption{The filling factor as a function of lag time $\tau$.
  It is evaluated on $(x(t), x(t-\tau), \dot{x}(t))$ phase space when $\Lambda=0$ (circles) and
$\Lambda=50$ (squares). We have taken the $10^2 \times 10^2 \times 10^2$ number of hypercubes for the region $x\in [0.0, 2.0]$
, $x(t-\tau) \in [0.0, 2.0]$, and $\dot{x} \in [-0.1, 0.1]$ and we have iterated the systems during $5 \times 10 ^4 $ seconds 
for each point.}
\end{center}
\end{figure}

Figure 3 shows the reconstructed phase trajectories 
for fixed and TDTs in $(x(t), x(t-\tau))$ space.
When the delay time is fixed (the first row of Fig. 3), 
the trajectory suddenly collapses into a quite simple shape  
at the value of $\tau=200$ (Fig. 3 (b)), while the others look very complex 
(we shall discuss the feature quantitively in the next paragraph).
This explains why the prediction error
sharply drops in the case of fixed delay time (filled triangles in Fig. 2).
On the contrary, when the delay time is time-dependent (the second row of Fig. 3), 
one can not see any qualitative difference which coincides with 
the fact that the prediction error is almost constant in Fig. 2 (filled circles). 
This is a genuine effect caused by the nature of TDT. 
It presents a possibility that the delayed system with TDT 
can be used in communication with better performance.     

To quantify the complexity of the attractors, we evaluate the filling factor
which is the normalized number of hypercubes visited by the projected trajectory \cite{Bunner}.
It is one of the measures which directly show the complexity of the projected 
attractor.  The results are presented in Fig. 4.
In the case of TDT (filled circles in Fig. 4),
the taken phase space is filled by $10\%$  and it hardly depends on the
chosen time lag $\tau$ for reconstruction which explains the above description for Fig. 3 (d) - (f). 
In case of fixed time delay, if the time lag appropriately chosen as $\tau=200$, the attractor
only fills $1\%$ of the taken phase space. 
Thus one may say that the reconstructed attractor of Fig. 2 (e)  
is 10 times as complex as that of  Fig. 2 (b) based the values of the filling factors.


\begin{figure}
\begin{center}
\rotatebox[origin=c]{0}{\includegraphics[width=8.3cm]{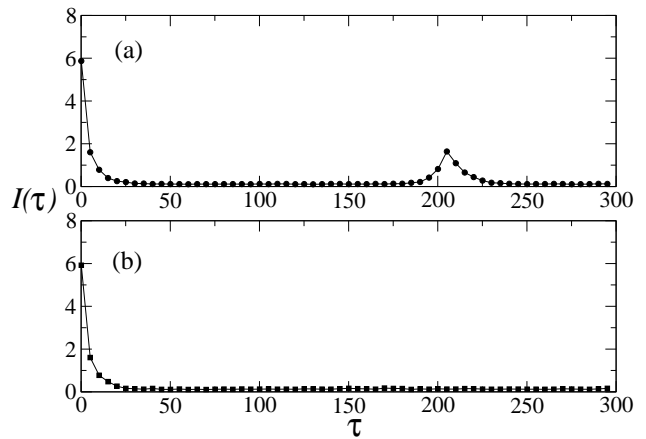}}
\caption{Average mutual information of the modified Mackey-Glass model: (a) fixed delay time ($\Lambda=0$) and 
(b) TDT at $\Lambda=270$.}
\end{center}
\end{figure}
In order to perform phase space reconstruction,
the first step one must take is to find out the lag time $\tau_e$ 
for the delayed coordinates \cite{Taken,Aba}.
It is usually determined by the first minimum point of the average mutual information.
The average mutual information is defined by \cite{Mut_Info}  
\begin{eqnarray}
I(\tau)&=&\sum _{x(n), x(n+\tau)} P(x(n), x(n+\tau)) \nonumber\\ 
        & &     \times \log _2 \left 
                [ \frac{P(x(n),x(n+\tau))}{P(x(n)) P(x(n+\tau))} \right].
\end{eqnarray}

Figure 5 shows the average mutual information in the cases of fixed delay 
and TDT.
For the fixed delay time, the average mutual information has a peak near $\tau=200$,
which has the same implication with that of the prediction error in Fig. 2. 
On the other hand, for TDT,
the average mutual information has the delta function shape.  
This means that the information of the observed time series 
deteriorates more rapidly by TDT, compared to that of the fixed delay.
Both of them have a wide range of degenerated minimum.
In the latter, one can expect that the delay coordinate for phase space reconstruction would not be so unique because  
the degenerated region is wider than that of the former. 


To analyze the global effects of the system according to the property of the driving signal, 
we consider the metric entropy defined by $h=\sum_{x(n)} P(x(n))\log_2[1/P(x(n))]$ which
is a measure of complexity or strength of nonlinearity of the system \cite{Map_Delay}. It 
corresponds to the value of the average mutual information $I(\tau)$ at $\tau=0$. 
Figure 6 shows the entropy as a function of $T$ and $\Lambda$.
The value of entropy increases as $\Lambda$ and $1/T$ both increase, 
while in the delayed system with a fixed delay time it is almost constant \cite{Map_Delay}.
Therefore, we understand that the profile of TDT plays a significant role 
that controls the complexity and nonlinearity of the time-delayed system.
All observations lead us to conclude that  
the nonlinearity characteristics of the time-delayed system are significantly 
changed depending on the properties of time-dependent delay time and, especially, 
that the reconstructed phase trajectory of the system 
is not collapsed into simple manifold, differently from the delayed system with fixed delay time.

\begin{figure}
\begin{center}
\rotatebox[origin=c]{0}{\includegraphics[width=8.2cm]{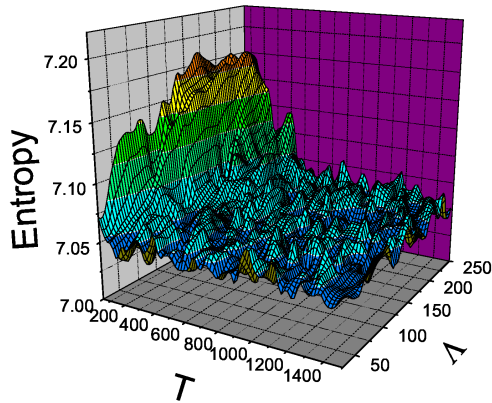}}
\caption{Entropy of the modified Mackey-Glass model 
as a function of $T$ and $\Lambda$}
\end{center}
\end{figure}

In conclusion, we have studied the characteristics of the time-delayed system in the 
presence of TDT.
By presenting the numerical evidence, 
we have shown that the time-delayed system with TDT transits 
to the uncollapsible hyperchaotic system.
This fact implies that the phase space reconstruction 
of the systems with TDT is hardly possible. 
The reason is that phase reconstruction methods
for a time-delayed system usually assume the exact determination 
of fixed delay time \cite{TD_Recon,Bunner}.  
We expect that these characteristics of the time-delayed system 
with TDT should be useful to implement communication systems. 


The authors thank M.-W. Kim and S.-Y. Lee  for helpful discussions.
This work is supported by Creative Research Initiatives 
of the Korean Ministry of Science and Technology.


\begin{thebibliography}{150}
\bibitem{Coeff} By applying the extremal conditions to the prediction error: 
                ${\partial \sigma_j^2}/{ \partial b_j}=0$ and 
                ${\partial \sigma_j^2}/{\partial \vec{a}_j}=0$,
                one obtain three equations:
                $b_j + a^0_j \langle x_i \rangle + a^1_j \langle x_{i-\tau} \rangle= \langle\hat{\dot{x}}_i \rangle$,
                $\langle x_i \rangle b_j + \langle x_i^2 \rangle a^0_j + \langle x_{i-\tau} x_i \rangle a^1_j= \langle
                \hat{\dot{x}}_i x_i\rangle$, 
                $\langle x_{i-\tau} \rangle b_j + \langle x_{i-\tau} x_i \rangle a^0_j + \langle x_{i-\tau}^2 \rangle a^1_j
                = \langle \hat{\dot{x}}_i x_{i-\tau}\rangle$,
                where $\langle X \rangle = \frac{1}{N_{U_j}} \sum_{\vec{v}_i \in U_j} X$.
                One can easily determine the parameters $b_j$ and $\vec{a}_j=(a^0_j, a^1_j)$ for least square fitting
                by solving above equations.
\end{thebibliography}
\end{document}